\documentstyle[11pt,fleqn]{article}  
\input{psfig.tex}                
\title{Possible Quantum Critical Region tuned by \\the Disorder and Pauli-Blocking Effects}
\author{A. Kwang-Hua Chu} 
\date{Department of Physics, Xinjiang University, \\
Urumqi 830046, PR China}
\begin{document}      
\maketitle
\begin{abstract}           
Based on the acoustic analog, we investigate both of the effects :
disorder and (Pauli-blocking) interaction to the possible
localization in electron gases by using the quantum discrete
kinetic model. Effects of the disorder (or free-orientation :
$\theta$ which is related to the relative direction of scattering
of particles w.r.t. to the normal of the propagating plane-wave
front) which is introduced into the \"{U}hling-Uhlenbeck equations
together with those of the Pauli-blocking  are presented. We
obtain the possible phase diagram (related to the strength of
disorders and the mean free path) which resembles qualitatively
that proposed by Abrahams$^{1}$.

\vspace{2mm} \noindent PACS : \hspace*{2mm}71.10.Hf; 71.30.+h;
73.43.Nq; 73.20.Jc; 73.20.Fz
\newline
\end{abstract}
\doublerulesep=6mm    
\baselineskip=6mm  
\bibliographystyle{plain}
An intriguing aspect of quantum mechanics is that even at absolute
zero temperature quantum fluctuations prevail in a system, whereas
all thermal fluctuations are frozen out. These quantum
fluctuations are able to induce a macroscopic phase transition in
the ground state of a many-body system, when the ratio of two
competing terms in the underlying Hamiltonian is varied across a
critical value [1-4]. The instability of the quantum critical
behavior with respect to (w.r.t.) the disorder can be interpreted
as a signal for phenomena of localization. Recently some of the
new developments in the localization problem \cite{QCP:2000} have
become a major theme in the condense matter research. One example
is the strongly-interacting electron (non-Fermi) liquid with
different strengths of disorder \cite{RMP:2001A}. Interesting
issues are the quantum phase transition (QPT) and quantum critical
point (QCP).  Both effects of disorder and interaction are closely
relevant to the weak and strong localization
\cite{Lat:Disord,MIT:JPC}. They are then related to the
metal-insulator transition in two-dimensions (2D). Although most
of theories proposed before are based on the Fermi liquid
behavior, new insights could be obtained considering the possible
analogy with superconducting transition which might be related to
the bosonic system [2-3]. \newline 
In the last two decades a variety of metals have been discovered
which display thermodynamic and transport properties at low
temperatures which are fundamentally different from those of the
usual metallic systems which are well described by the Landau
Fermi-liquid theory. The resistivity in a variety of high mobility
2D electron/hole systems is seen experimentally to exhibit a
number of interesting anomalies that don't as yet have an adequate
theoretical understanding.
Note that, at sufficiently low electron densities, an ideal
two-dimensional electron systems becomes strongly correlated,
because the kinetic energy is overpowered by energy of
electron-electron interactions (exchange and correlation energy).
The interaction strength is normally described by the Wigner-Seitz
radius, $r_s =1/(\pi n_s)^{1/2} a_B$ (where $n_s$ is the electron
density and $a_B$ is the effective Bohr radius in semiconductor).
%
Till now, for rather large $r_s$ ($\gg 1$) and together with
disorder, the nature of this metal-insulator transition (MIT)
remains the subject of ongoing debate [6,8-9]. 
\newline Motivated by the analogy
between electrons in periodic or disordered metals and waves in
classical acoustical systems [10-12] an investigation for
observing possible QCP in  MIT or relevant localization
\cite{Lat:Disord} using the quantum discrete kinetic model [13-14]
was performed and will be presented here. In present approach the
\"{U}hling-Uhlenbeck collision term [13] which could describe the
collision of a gas of dilute hard-sphere Fermi- or Bose-particles
by tuning a parameter $\gamma$ : a Pauli-blocking factor (or
$\gamma f$ with $f$ being a normalized (continuous) distribution
function giving the number of particles per cell) is adopted
together with a disorder or free-orientation ($\theta$ which is
related to the relative direction of scattering of particles
w.r.t. to the normal of the propagating plane-wave front) into the
quantum discrete kinetic model which can be used to obtain
dispersion relations of (plane) sound waves propagating in quantum
gases. We then study the quantum critical behavior based on the
acoustical analog [15-17] which has been verified before. The
possible phase diagram for MIT  and/or resistance(or
disorder)-scattering amplitude curves (as the temperature is
decreased ) we obtained resemble qualitatively those proposed in
[5] (cf Fig. 4 therein). The non-Fermi liquid behavior was also
clearly illustrated here.
\newline We firstly introduce the concept of acoustical
analog [15] in brief.
In a mesoscopic system, where the sample size is smaller than the
mean free path for an elastic scattering, it is satisfactory for a
one-electron model to solve the time-independent Schr\"{o}dinger
equation :
 $-({\hbar^2}/{2m}) \nabla^2 \psi + V' (\vec{r}) \psi = E \psi$
or (after dividing by $-\hbar^2/2m$)
 $\nabla^2 \psi + [q^2 - V (\vec{r})] \psi = 0$,
where $q$ is an (energy) eigenvalue parameter, which for the
quantum-mechanic system is $\sqrt{2mE/\hbar^2}$. Meanwhile, the
equation for classical (scalar) waves is
 $\nabla^2 \psi - ({\partial^2 \psi}/{c^2 \,\partial t^2})
 =0$
or (after applying a Fourier transform in time and contriving a
system where $c$ (the wave speed) varies with position $\vec{r}$)
 $\nabla^2 \psi + [q^2 - V (\vec{r})] \psi = 0$,
here, the eigenvalue parameter $q$ is $\omega/c_0$, where $\omega$
is a natural frequency and $c_0$ is a reference wave speed.
Comparing the time dependencies one gets the quantum and classical
relation $E= \hbar \omega$ [15-17].\newline 
We assume that the gas is composed of identical hard-sphere
particles of the same mass \cite{U:U,Chu:PhD,Platkowski:1988}. The
velocities of these particles are restricted to, e.g., : ${\bf
u}_1, {\bf u}_2, \cdots, {\bf u}_p$, $p$ is a finite positive
integer. The discrete number densities of particles are denoted by
$N_i ({\bf x},t)$ associated with the velocities ${\bf u}_i$ at
point ${\bf x}$ and time $t$. If only nonlinear binary collisions
and the evolution of $N_i$ are considered, we have
\begin{equation}
 \frac{\partial N_i}{\partial t}+ {\bf u}_i \cdot \nabla N_i
 = F_i \equiv \frac{1}{2}\sum_{j,k,l} (A^{ij}_{kl} N_k N_l - A_{ij}^{kl}
 N_i N_j),
\hspace*{3mm} i \in\Lambda =\{1,\cdots,p\},
\end{equation}
where $(i,j)$ and $(k,l)$ ($i\not=j$ or $k\not=l$) are admissible
sets of collisions [13-14,16-18]
Here, the summation is taken over all $j,k,l \in \Lambda$, where
$A_{kl}^{ij}$ are nonnegative constants satisfying [13-14,18]
 $ A_{kl}^{ji}=A_{kl}^{ij}=A_{lk}^{ij}$, 
 $ A_{kl}^{ij} ({\bf u}_i +{\bf u}_j -{\bf u}_k -{\bf u}_l )=0$,
 and $A_{kl}^{ij}=A_{ij}^{kl}$. 
The conditions defined for the discrete velocities above require
that there are elastic, binary collisions, such that momentum and
energy are preserved, i.e.,
 ${\bf u}_i +{\bf u}_j = {\bf u}_k +{\bf u}_l$,
 $|{\bf u}_i|^2 +|{\bf u}_j|^2 = |{\bf u}_k|^2 +|{\bf u}_l|^2$,
are possible for $1\le i,j,k,l\le p$.  
We note that, the summation of $N_i$ ($\sum_i N_i$) : the total
discrete number density here is related to the macroscopic density
: $\rho \,(= m_p \sum_i N_i)$, where $m_p$ is the mass of the
particle \cite{Platkowski:1988}.
\newline
Together with the introducing of the \"{U}hling-Uhlenbeck
collision term \cite{U:U} :
 $F_i$ $=\sum_{j,k,l} A^{ij}_{kl} \,[ N_k N_l$ $(1+\gamma N_i)(1+\gamma N_j)$ $-
 N_i N_j (1+\gamma N_k)(1+\gamma N_l)]$,
into equation (1), for $\gamma <0$ (normally, $\gamma=-1$), we can
then obtain a quantum discrete kinetic equation for a gas of
Fermi-particles; while for $\gamma
> 0$ (normally, $\gamma=1$) we obtain one for a gas of Bose-particles,
and for $\gamma =0$ we recover the equation (1).  \newline
Considering binary  collisions only, from equation above, the
model of quantum discrete kinetic equation for Fermi or Bose gases
proposed in [13-14] is then a system of $2n(=p)$ semilinear
partial differential equations of the hyperbolic type :
\begin{displaymath}
 \frac{\partial}{\partial t}N_i +{\bf v}_i \cdot\frac{\partial}{\partial
 {\bf x}} N_i =\frac{c S}{n} \sum_{j=1}^{2n} N_j N_{j+n}(1+\gamma N_{j+1})
 (1+\gamma N_{j+n+1})-
\end{displaymath}
\begin{equation}
 \hspace*{18mm} 2 c S N_i  N_{i+n} (1+\gamma N_{i+1})(1+\gamma
 N_{i+n+1}),\hspace*{24mm} i=1,\cdots, 2 n,
\end{equation}
where $N_i=N_{i+2n}$ are unknown functions, and ${\bf v}_i$ =$ c
(\cos[\theta+(i-1) \pi/n], \sin[\theta+(i-1)\pi/n])$; $c$ is a
reference velocity modulus and the same order of magnitude as that
($c$, the sound speed in the absence of scatters) used in Ref. 9,
$\theta$ is the orientation starting from the positive $x-$axis to
the $u_1$ direction and could be thought of as a parameter for
introducing a {\it disorder}
\cite{RMP:Loc2001,Local:1996,Chu:2001,ChuA:2001,ChuA:2002A}, $S$
is an effective collision cross-section for the collision system.
\newline Since passage of the sound wave will cause a small departure
from an equilibrium state and result in energy loss owing to
internal friction and heat conduction, we linearize above
equations around a uniform equilibrium state (particles' number
density : $N_0$) by setting $N_i (t,x)$ =$N_0$ $(1+P_i (t,x))$,
where $P_i$ is a small perturbation. 
After some similar manipulations as mentioned in [14-17], with
$B=\gamma N_0 <0$ \cite{U:U,Chu:PhD}, which gives or defines the
(proportional) contribution from the Fermi gases (if $\gamma < 0$,
e.g., $\gamma=-1$), we then have
\begin{equation}
 [\frac{\partial^2 }{\partial t^2} +c^2
 \cos^2[\theta+\frac{(m-1)\pi}{n}]
 \frac{\partial^2 }{\partial x^2} +4 c S N_0 (1+B) \frac{\partial
 }{\partial t}] D_m= \frac{4 c S N_0 (1+B)}{n} \sum_{k=1}^{n} \frac{\partial
 }{\partial t} D_k  ,
\end{equation}
where $D_m =(P_m +P_{m+n})/2$, $m=1,\cdots,n$, since $D_1 =D_m$
for $1=m$ (mod $2 n)$. \newline 
We are ready to look for the solutions in the form of plane wave
$D_m$= $a_m$ exp $i (k x- \omega t)$, $(m=1,\cdots,n)$, with
$\omega$=$\omega(k)$. This is related to the dispersion relations
of 1D (forced) plane wave propagation in Fermi gases. So we have
\begin{equation}
 (1+i h (1+B)-2 \lambda^2 cos^2 [\theta+\frac{(m-1)\pi}{n}]) a_m -\frac{i h (1+B)}{n}
 \sum_{k=1}^n a_k =0  , \hspace*{6mm} m=1,\cdots,n,
\end{equation}
where
\begin{displaymath}
\lambda=k c/(\sqrt{2}\omega),  \hspace*{18mm} h=4 c S N_0 /\omega
\hspace*{6mm} \propto \hspace*{2mm} 1/K_n,
\end{displaymath}
where $h$ is the rarefaction parameter of the gas; $K_n$ is the
Knudsen number which is defined as the ratio of the mean free path
of gases to the wave length of the plane (sound) wave.
\newline
We can obtain the complex spectra ($\lambda=\lambda_r +$ i
$\lambda_i$; $\lambda_r = k_r c/(\sqrt{2}\omega)$: sound
dispersion, a relative measure of the sound or phase speed;
$\lambda_i = k_i c/(\sqrt{2}\omega)$ : sound attenuation or
absorption) from the complex polynomial equation above. Here, $B$
could be related to the occupation number of different-statistic
particles of gases
To examine the critical region possibly tuned by the
Pauli-blocking measure $B=\gamma N_0$ and the disorder $\theta$,
as evidenced from our preliminary results : $\lambda_i =0$ for
cases of $B=-1$ or $\theta=\pi/4$ [17], we firstly check those
spectra near $\theta=0$, say, $\theta=0.005$ and $\theta=\pi/4
\approx 0.7854$, say, $\theta=0.78535$ for a $B$-sweep ($B$
decreases from 1 to -1). We plot them into figures 1, 2,
respectively. Note that, as the disorder or free-orientation
$\theta$ is not zero, there will be two kinds of propagation of
the disturbance wave : sound and diffusion modes [16-17,19-20].
The latter (anomalous) mode has been reported in Boltzmann gases
[19,21] and is related to the propagation of entropy wave which is
not used in the acoustical analog here. The absence of (further)
diffusion (or maximum absorption) for the sound mode at certain
state ($h$, corresponding to the inverse of energy $E$; cf. Refs.
11 or 15) is classified as a localized state [11,15,17]. The state
of decreasing $h$ corresponds to that of $T$ (absolute
temperature) decreasing as the mean free path is increasing
(density or pressure decreasing).
\newline We can observe the max. $\lambda_i$ (absorption of sound
mode, relevant to the localization length according to the
acoustical analog [15,17]) drop to around four orders of magnitude
from $\theta=0.005$ to $0.78535$! This is a clear demonstration of
the effect of disorder. Meanwhile, once the Pauli-blocking measure
($B$) increases or decreases from zero (Boltzmann gases), the
latter (Fermi gases : $B <0$) shows opposite trend compared to
that of the former (Bose gases : $B>0$) considering the shift of
the max. $\lambda_i$ state ($\delta h$). $\delta h >0$ is  for
Fermi gases ($|B|$ increasing), and  the reverse ($\delta h <0$)
is for Bose gases ($B$ increasing)! This illustrates partly the
electron-electron interaction effect (through the Pauli exclusion
principle). These results will be crucial for further obtaining
the phase diagram (flow to metallic or insulating state as the
density or temperature is decreased) tuned by both disorder and
the (electron-electron) interaction below. Here, $B=-1$ or
$\theta=0, \pi/4$ might be fixed points commented in [22].
\newline
To check what happens
when the temperature is decreased (or the density or $h$ is
decreased) to near $T=0$, we collect all the data based on the
acoustical analog from the dispersion relations (especially the
absorption of sound mode) we calculated for ranges in different
degrees of disorder (here, $\theta$ is up to $\pi/4$ considering
single-particle scattering and binary collisions; in fact, effects
of $\theta$ are symmetric w.r.t. $\theta=\pi/4$ for
$0\le\theta\le\pi/2$ [17]) and Pauli-blocking measure. After that,
we plot the possible phase diagram for the (dimensionless)
conductivity vs. the absolute (dimensionless) temperature into
Fig. 3 (for different $B$s : $B=-0.9, -0.7, -0.5, -0.3, 0.1$).
Here, MFP is
the mean free path and the temperature vs. MFP relations could be
traced  from [23] (cf. Fig. 3 therein). Note that, the resistivity
or resistance is proportional to the strength of disorder in 2D
(in the sense that for weak disorder it is given by $1/(k_F l)$,
in units of $\hbar/e^2$; $k_F$ is the Fermi wave number, $l$ is
the mean free path associated with the usual Drude conductivity)
[5]. This figure shows that as the temperature decreases to a
rather low value, the resistivity (or strength of disorder) will
decrease sharply (at least for Bose or Fermi gases). There is no
doubt that this result resembles qualitatively that proposed
before [5,24-25].
\newline
To know the detailed effects of electron-electron interactions
(tuned by $B$s here), we plot the corresponding figure as shown in
Fig. 4. Each contour line (flow path relevant to a specific $B$
tuned by the disorder) represents the behavior when $T$ is
decreased, and different (flow) lines represent different values
of scattering amplitude ($S_b$ $\propto K_n$ or $r_s$).
Interesting results are (i) there seems to be a quantum phase
transition boundary (interface or regime) for bosonic-like
particles ($B>0$) which resembles that of high-temperature
superconducting phase transition due to doping if we treat the
strength of disorder to be equivalent to the doping amount! (ii)
as evidenced in the top part of this plot for the Bose gases (near
QCPs for smaller resistivity together with rather large arrows),
it confirms Larkin's comment (cf page 793 in [2] by Larkin) : both
Bose and Fermi approached are important for the (QP) transition;
Bose approach is more useful in a small region close to the
transition!
\newline To make sure we already recover previous proposed results
(possible renormalization group (RG) flows for disorder plus
interactions or scattering amplitude-resistance curves or
suggested phase diagram tuned by $r_s$ and disorder, cf. [5] or
[24-25]), we summarize our results by illustrating them into a 3D
plot as demonstrated in Fig. 5. Crossover lines separate those
flows which begin at $T>0$ at small resistance (disorder  or the
inverse of conductance : $1/G$) and flow initially, as the
temperature is lowered, toward larger resistance but then are
repelled by the fixed point and flow to large $S_b$ (or $r_s$) and
large G (metallic behavior) from those which begin at larger
disorder (or smaller G) and flow toward very-large disorder
(insulating behavior). Again, this result resembles that proposed
before (cf. [5] or [24-25] therein). Note that, this flow is also
similar to those in [24-25] (e.g., cf. Fig. 41 by Aoki [24]).
Meanwhile, as expected before, at $T=0$, a 2D system would become
a Wigner crystal phase at $r_s \ge 37$ [26]. This lies possibly
near or above the rather large $S_b$ position with zero strength
of disorder in our illustrations. Possible QCPs happen around $S_b
\sim 80$ for $B=0.85$ but $S_b \sim 10$ for $B=0.1$.
\newline To conclude in brief, our illustrations here, although
are based on the acoustical analog  of our quantum discrete
kinetic calculations, can indeed show the non-Fermi liquid and
quantum critical behavior for the metal-insulator transition in 2D
[27-31] as the temperature is decreased to rather low values.

%

\newpage

\psfig{file=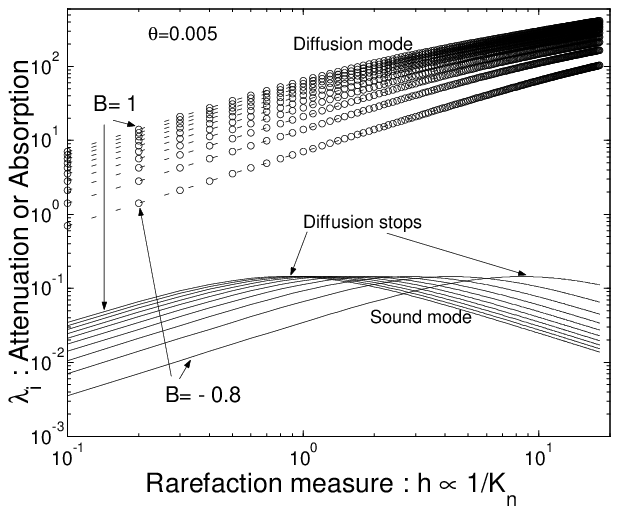,bbllx=0.0cm,bblly=18cm,bburx=10cm,bbury=28cm,rheight=8cm,rwidth=8cm,clip=}

\begin{figure}[h]

\hspace*{12mm} Fig. 1 \hspace*{3mm}  {\small Variations of the
smaller (Sound mode) and larger (Diffusion mode [19,21])
\newline \hspace*{12mm} $\lambda_i$
w.r.t. $h$ for the same disorder : $\theta=0.005$ in different
Pauli-blocking measures \newline \hspace*{12mm} ($B>0$ corresponds
to Bose gases). The maximum absorption states for sound modes
\newline \hspace*{12mm} correspond to possible localized states  as the diffusion stops.
$B$ decreases from $1$
\newline \hspace*{10mm}  (the highest)  to
$-1$ with the decrement being $-0.2$. $\lambda_i =0$ as $B=-1$. $h
=4 c S N_0/\omega$. \newline \hspace*{10mm} $N_0$ is the number density
and $S$ is the effective collision cross-section.}
\end{figure}

\psfig{file=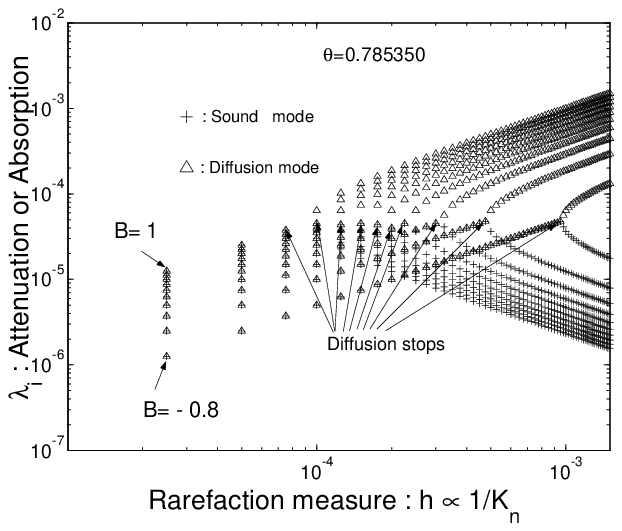,bbllx=0.0cm,bblly=18cm,bburx=10cm,bbury=28cm,rheight=8cm,rwidth=8cm,clip=}

\begin{figure}[h]
\hspace*{10mm} Fig. 2 \hspace*{1mm} {\small Pauli-blocking ($B$)
effects on the absorption $\lambda_i$. (Note that the \newline
\hspace*{10mm} localization length $\propto 1/\lambda_i$, cf.
[11,16-17]. The energy $E$
 (relevant to the \newline \hspace*{10mm} illustration of localized states
in [11]) corresponds to $\hbar \omega$ [15-16]. $E \propto 1/h$
\newline \hspace*{10mm}  once $c\,S\,N_0$ is fixed.) $B$ decreases
from $1$ (the highest)  to $-1$ with the
\newline \hspace*{10mm} decrement  being $-0.2$. Bose and Fermi
gases are in opposite trend for
\newline \hspace*{10mm}   the same
increment $|B|=0.2$. $\theta=0.78535$ and $\lambda_i=0$ as
$\theta=\pi/4 \approx 0.7854$.}
\end{figure}

\newpage

\psfig{file=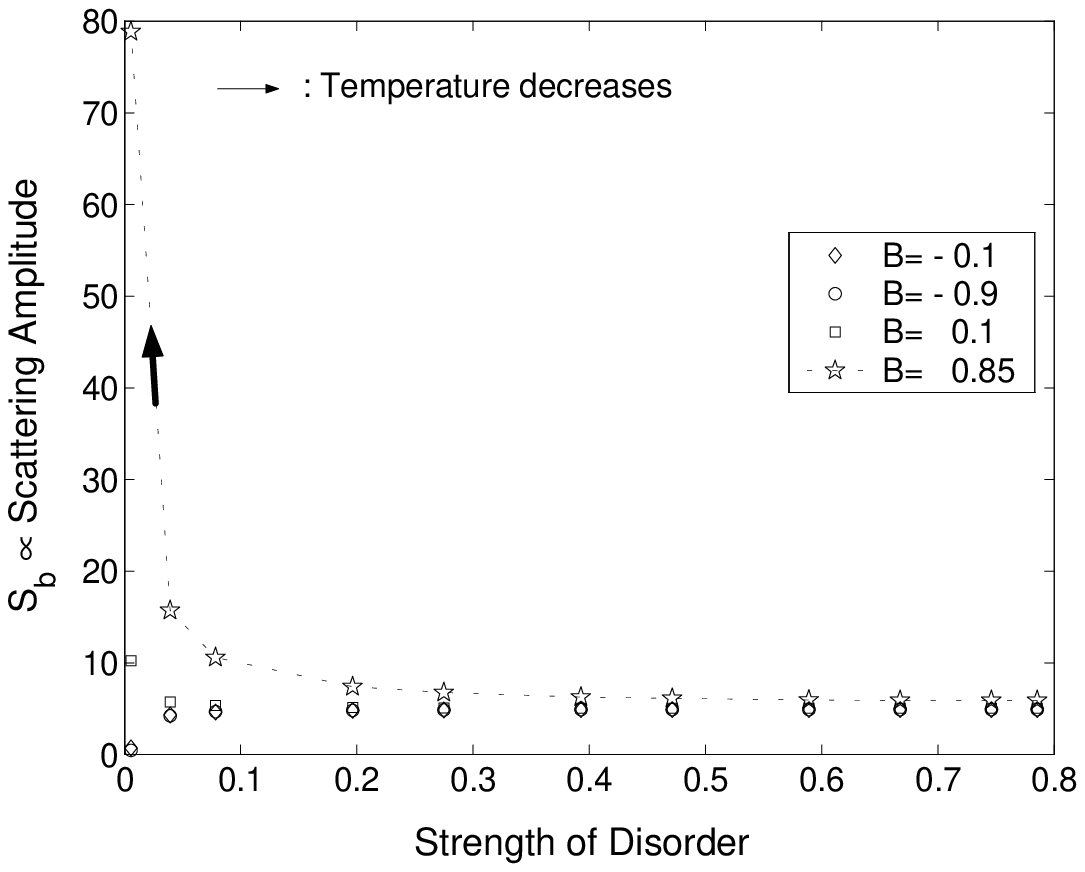,bbllx=0.0cm,bblly=12.8cm,bburx=14cm,bbury=24.2cm,rheight=9.6cm,rwidth=9.6cm,clip=}

\begin{figure}[h]
\hspace*{10mm} Fig. 3 \hspace*{1mm} Possible phase diagram for
non-Fermi gases w.r.t. the disorder and \newline \hspace*{10mm}
the scattering amplitude $S_b$. $S_b \propto  B K_n$ and $K_n
\propto$ the mean free path. Note that  \newline \hspace*{10mm}
the resistivity is proportional to  the strength of disorder in 2D
[5-6]. Here, once
\newline \hspace*{10mm}     the
temperature is  lowered down the resistivity firstly increases and
 then decreases.
\end{figure}

\psfig{file=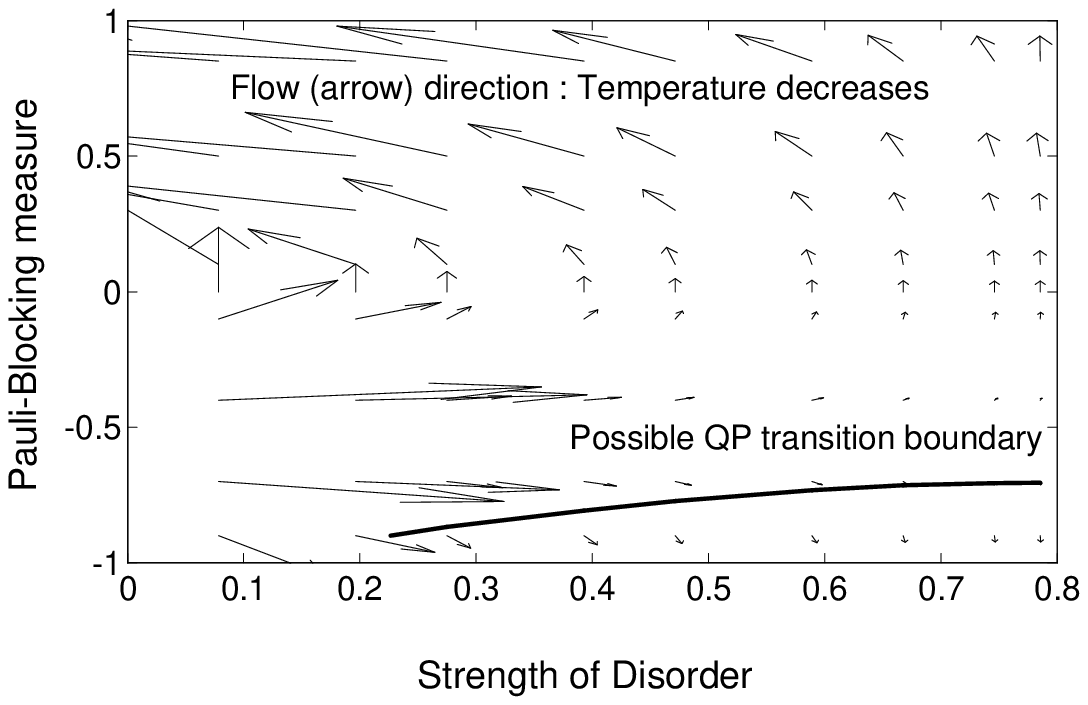,bbllx=0.0cm,bblly=13cm,bburx=14cm,bbury=24.2cm,rheight=9.5cm,rwidth=9.5cm,clip=}

\begin{figure}[h]
\hspace*{10mm} Fig. 4 \hspace*{1mm} Possible flow pattern for
$S_b$ w.r.t. the disorder and Pauli-blocking \newline
\hspace*{10mm}
 measure ($B$). $S_b \propto  B K_n$ and $K_n
\propto$ the mean free path. Note that  the \newline
\hspace*{10mm} resistivity is proportional to  the disorder [5-6]
in 2D. Each continuous flow
\newline \hspace*{10mm} represents the behavior when $T$ is
decreased, and different paths represent \newline \hspace*{10mm}
different values of Pauli-blocking measure. (cf Fig. 4 in [5])
\end{figure}

\newpage

\psfig{file=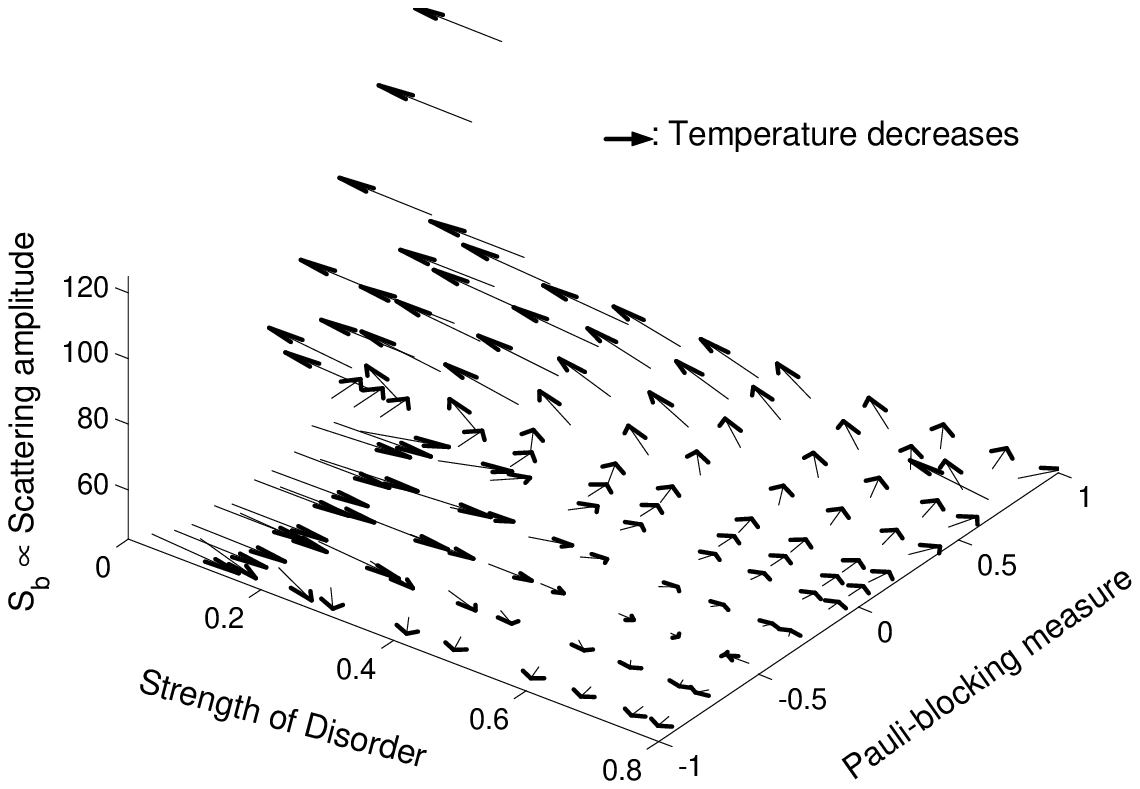,bbllx=0.0cm,bblly=12.8cm,bburx=14cm,bbury=24.2cm,rheight=10cm,rwidth=10cm,clip=}

\begin{figure}[h]
\hspace*{10mm} Fig. 5 \hspace*{1mm} Possible scattering amplitude
($S_b$) vs. disorder and Pauli-blocking \newline \hspace*{10mm}
measure curves. $S_b \propto B K_n$, and $K_n \propto$ the mean
free path. Note that \newline \hspace*{10mm} the resistivity  is
proportional to the strength of disorder in 2D [6].  Here,
\newline \hspace*{10mm} the
flow (arrow) direction means that the temperature is decreasing.
\newline \hspace*{10mm}
These results resemble qualitatively those proposed by Abrahams in
[5-6].
\end{figure}
\end{document}